\documentclass[twocolumn]{aastex63}
\usepackage[utf8]{inputenc}
\usepackage{comment}
\usepackage{amsmath}
\usepackage{graphicx}
\usepackage{multirow}
\usepackage{booktabs}
\usepackage{tabularx}

\usepackage{lineno}

\begin{document}

\title{A power spectral study of PHANGS galaxies with JWST MIRI: \\ On the spatial scales of dust and PAHs}


\author{Charlie Lind-Thomsen}
\affiliation{Cosmic Dawn Center (DAWN)}
\affiliation{Niels Bohr Institute, University of Copenhagen, Jagtvej 128, DK-2200, Copenhagen N, Denmark}
\author[0000-0002-5460-6126]{Albert Sneppen}
\affiliation{Cosmic Dawn Center (DAWN)}
\affiliation{Niels Bohr Institute, University of Copenhagen, Jagtvej 128, DK-2200, Copenhagen N, Denmark}
\author[0000-0003-3780-6801]{Charles L. Steinhardt}
\affiliation{Department of Physics and Astronomy, University of Missouri, Columbia, MO 65211, USA}
\affiliation{DARK Cosmology Center, University of Copenhagen, Jagtvej 155A, DK-2200, Copenhagen N, Denmark}
\affiliation{Cosmic Dawn Center (DAWN)}

\begin{abstract}
The interstellar medium (ISM) consists of a diversity of structures across a range of spatial scales, intimately tied to galactic evolution. In this work, Fourier analysis is used to characterize the spatial structures of dust and Polycyclic Aromatic Hydrocarbons (PAHs) in the ISM of PHANGS-JWST galaxies, observed in the four photometric mid-infrared (MIR) filters from F770W to F2100W (i.e., 7.7 to 21$\mu$m)). We quantify the abundance of structure on different spatial scales using the power-law slope, $\alpha$, of the spatial power spectra. The distribution of $\alpha$ across all length scales differs significantly between filters, with steeper slopes for PAH-dominated filters ($\alpha_{F770W} = 2.19^{+0.16}_{-0.15}$, $\alpha_{F1130W} = 1.88^{+0.25}_{-0.37}$) and shallower for the dust-continuum ($\alpha_{F1000W} = 1.48^{+0.33}_{-0.47}$, $\alpha_{F2100W} = 0.94^{+0.23}_{-0.28}$).
The distribution of $\alpha$ across galaxies is narrower for PAH-dominated than for thermal dust-dominated bands, highlighting that PAHs trace photo-dissociation regions dominated by similar physical processes, whereas dust structures are an integrated property over the diverse evolutionary histories of their host galaxies. Unlike dust structures, PAH-sensitive bands display a break in the power spectrum: below a characteristic scale, $\ell_0=160\mathrm{pc}^{+110\mathrm{pc}}_{-50\mathrm{pc}}$, PAH structures are suppressed.
\newline


\end{abstract}

\section{Introduction}\label{sec:Intro}

\begin{figure*}[t!]
    \centering
    \includegraphics[width=\linewidth]{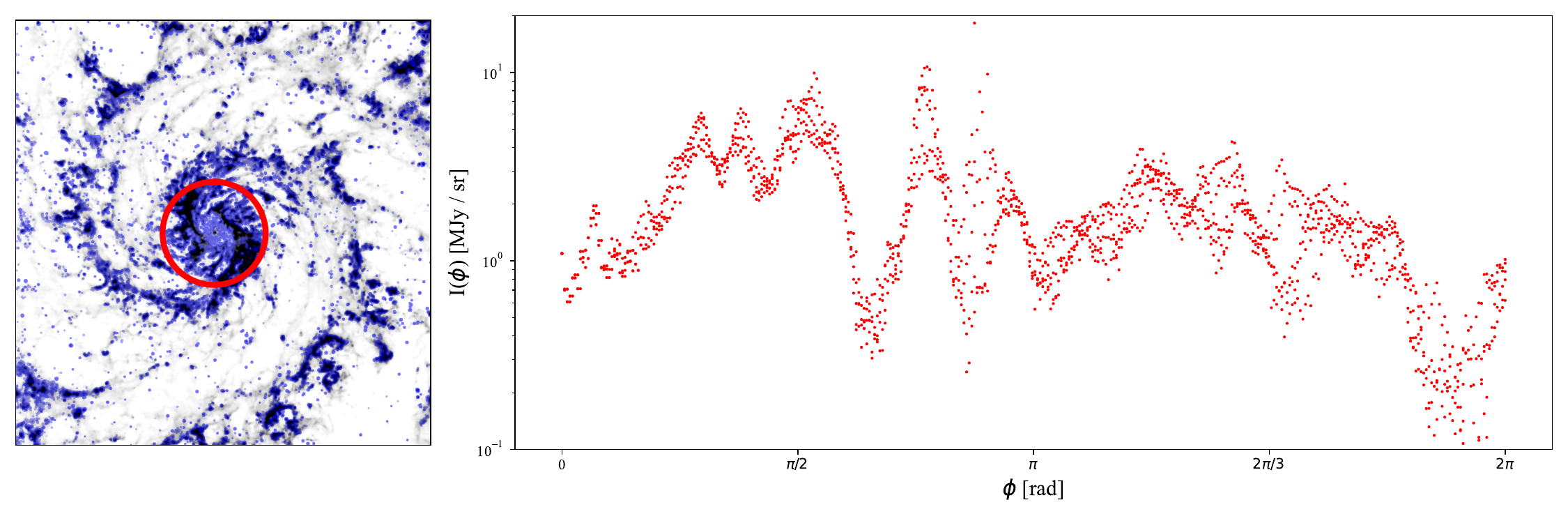}
    \caption{(left) Imaging of NGC 628 from the PHANGS-JWST survey in F770W, a photometric filter dominated by PAH emission at 7.7 $\mu\text{m}$ overlaid with contours of F1000W imaging, a photometric filter dominated by the dust continuum emission. The red ring correspond to a radius of $r = 584 \text{pc} \pm 11 \text{pc}$ in the galaxy. (right) Intensity of NGC 628 at F770W at the radius $r$.The azimuthal fluctuations show a wave-like structure. This motivates Fourier analysis as a data processing method to study galaxy structure. From the contour of F1000W we see that dust and PAH have the same radial extent on the scales visually prominent on galactic scales,  ($\sim$ kpc size). }
    \label{fig:RadialSlice}
\end{figure*}
The interstellar medium (ISM) is intrinsically linked to galactic evolution \citep[e.g.][]{Heiles_2019}; thus, understanding its spatial structure is essential for studying galactic dynamics and star formation. The ISM exhibits structure across a wide range of spatial scales, from large-scale spiral arms and galactic disks to small-scale gas clouds and filaments.
It is shaped by physical processes spanning many orders of magnitude in both energy and scale $-$ from large-scale gravitational instabilities, spiral density waves, and shear instabilities, to superbubbles \citep{Kornreich_2000, Chu2000}, down to small-scale effects such as supernovae, massive star winds, and H II regions \citep{Bieging_1990}, with influences extending even to AU-scale phenomena like stellar winds.
Turbulence driven by supernovae (SNe) reproduces many observed characteristics of the ISM, such as the velocity dispersion of H I gas in galactic disks and the transition to starburst regimes \citep{1999_Korpi, 2006_Dib, Mee_2006, 2010_Bournaud}, underscoring the critical role of SNe feedback.
The high complexity of the ISM makes it challenging to study theoretically, and the relative contributions of different physical processes throughout the full extent of galactic disks are still not fully understood. Therefore, quantifying ISM morphology from imaging offers valuable insights into the nature of its structure and the dominant physical mechanisms at play.

The morphology of galaxies has been studied through the power spectrum in the Fourier domain, both of the turbulent velocity field and intensity, to characterize turbulence in galaxies for over 50 years \citep{kalnajs1975dynamique, Stanimirovic_1999, Lazarian_2000, Elmegreen_2001, Elmegreen2003, Elmegreen2003b, 2008_Dib, Block2009}.
Astrophysical structures have also been studied using other types of correlation functions, such as the $\Delta$-variance spectrum \citep{Elmegreen_2001, Dib_2020, 2021_Dib}, which is more sensitive to deviations from self-similarity than the power spectrum.
Previous studies using the power spectrum have shown that the spatial distribution of emission in galaxies, over various ranges of scales, can often be well described by a power law.
Within a range where a single power law is applicable, the slope $\alpha$ carries information about the relative abundance of structure across self-similar scales—this is the focus of our work. However, across broader ranges, the power spectrum is not always described by a single slope.
\cite{2021_Dib}, for example, show that the power spectra of 33 galaxies in The HI Nearby Galaxy Survey (THINGS) follow two distinct power-law regimes. They interpret this as a result of different temperature phases of the HI gas dominating at different scales, with a shallower slope corresponding to colder HI and a steeper slope to warmer HI.

Power spectra of turbulent media are generally expected to follow a single power law, with the slope depending on the properties of the turbulence (e.g., whether it is subsonic, and on its dimensionality). Accordingly, observed slopes have been found to align with theoretical expectations for different turbulence regimes.

A wide range of galactic objects have been studied through Fourier analysis. \cite{Stanimirovic_1999} examined the Small Magellanic Cloud (SMC) using high-resolution HI radio observations from the Australia Telescope Compact Array (ATCA), finding a slope of $\alpha \approx 3$ over nearly two decades of spatial scale ($10^2$–$10^4$ pc).
Similarly, \cite{Elmegreen_2001} used ATCA to study HI regions in the Large Magellanic Cloud (LMC), finding $\alpha \approx 8/3$ in a 2D power spectrum.
Attempts to extend this analysis into the optical, using the {\it Hubble Space Telescope} (HST) \citep{Elmegreen2003}, and into the mid-infrared (MIR), using the {\it Spitzer Space Telescope} \citep{Block2009}, have also been made. These studies determined power spectrum slopes of $\alpha \approx 3$ in the optical (based on six galaxies) and $\alpha \approx 2$ in the MIR (based on 33 nearby galaxies).

The varying slopes have all been interpreted as consistent with a highly turbulent ISM \citep[e.g.,][]{Falceta_Gon_alves_2014}, where the differences reflect variations in the physical properties of the turbulence, such as velocity and dimensionality.

In this work, we study the highly wavelength-dependent structure of the ISM in the MIR, and therefore interpret the power spectrum of galaxies outside the typical framework of turbulence. Instead, we examine how structure and morphology can be quantified through the Fourier domain.
The unprecedented resolution in the MIR provided by the {\it James Webb Space Telescope} (JWST) enables us to characterize galaxy morphology in the Fourier domain across $\sim$3 decades of spatial scale ($\sim10^4$–20 pc). Although turbulence is present in the ISM, we show that the morphology of structure, as described in the Fourier domain, is closely connected to the constituents traced by the specific photometric band. Thus, the power spectrum carries information about both the similarities and differences in the structural distribution of various ISM components.

In Sect.\ref{sec:Method}, we present the galaxies from the Physics at High Angular Resolution in Nearby GalaxieS (PHANGS)–JWST survey used in this study, and introduce the Fourier transform methodology and data processing steps. In Sect.\ref{sec:result}, we determine the power spectra in four MIR filters and present the corresponding slopes over different scale ranges, along with a measurement of the break scale for PAH structures, $\ell_0$. In Sect.~\ref{sec:discuss}, we argue that filters tracing PAHs are more similar to one another than to the filter tracing thermal dust, and that small-scale structure in the photometric bands dominated by PAHs is generally more suppressed than in bands dominated by thermal dust. Furthermore, PAH structures appear to have a characteristic scale, in contrast to dust structures, which show no such scale.

\section{Methodology} \label{sec:Method}

\subsection{Observations}

\begin{figure*}[t]
    \centering
    \includegraphics[width=\linewidth]{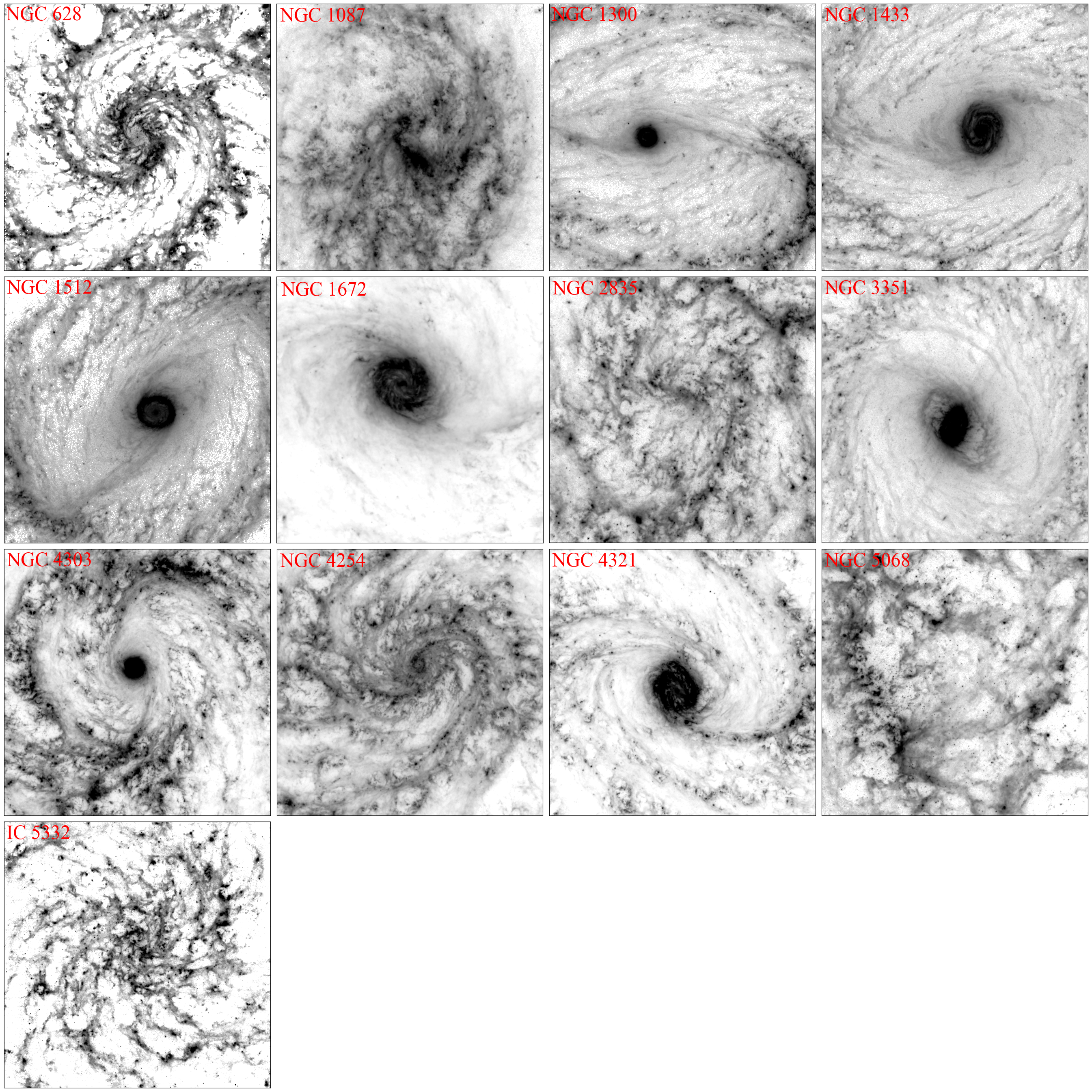}
    \caption{ Imaging of 13 galaxies (NGC\,628, NGC\,1087, NGC\,1300, NGC\,1433, NGC\,1512, NGC\,1672, NGC\,2835, NGC\,3351, NGC\,4254, NGC\,4303, NGC\,4321, NGC\,5068, IC\,5332) in the F770W photometric band from the PHANGS-JWST \citep{Phangs2022} galaxy survey on a logarithmic color intensity scale. Each galaxy has different upper and lower bounds on the colorbar, as this makes it possible to display the structure within each galaxy. The studied galaxies show a diversity in morphological properties, making it possible to study both universal and morphology specific aspects of the Fourier domain of galaxies. } \label{fig:Galaxies at F770W}
\end{figure*}

In this work, imaging of 13 galaxies from the PHANGS-JWST treasury \citep{Phangs2022} (NGC,628, NGC,1087, NGC,1300, NGC,1433, NGC,1512, NGC,1672, NGC,2835, NGC,3351, NGC,4254, NGC,4303, NGC,4321, NGC,5068, IC,5332) is studied. The imaging was obtained using the Mid-Infrared Imager Module (MIRIM) \citep{Bouchet_2015}, which includes the filters F770W, F1000W, F1130W and F2100W (see Figure~\ref{fig:Galaxies at F770W}). The imaging were processed as described in \cite{2024_Williams}. Specifications for the photometric bands are listed in Table~\ref{tab:FWHM}.

Five galaxies from the PHANGS-JWST treasury (NGC\,1365, NGC\,1566, NGC\,3627, NGC\,4535, and NGC\,7496) are excluded from this study due to diffraction spikes in the imaging, which introduce artificial scales in the Fourier domain.
Plotting an azimuthal slice of the imaging intensity (see Figure~\ref{fig:RadialSlice}), a superposition of wavelike structures with several different characteristic frequencies is seen, motivating the use of Fourier analysis to study the structure of galaxies.

\begin{table}[b]
\caption{Pixel and angular resolution for the MIRI photometric bands \citep{Phangs2022}. Bandwidth is the width of the filters, and the central wavelength is the place of the expected peak for emission of galaxies. F770W and F1130W are expected to trace PAH emission, while F1000W and F2100W are expected to trace the dust continuum emission. }
\begin{tabularx}{\columnwidth}{XXXXX}
\toprule[1pt] 
\toprule[1pt] 
Filter &  FWHM & FWHM & Band- width & Central wavelength\\
 & (pixel) & (arcsec) & $\mu$m & $\mu$m \\
\specialrule{0.1em}{0em}{0em} 
F770W & 2.27 & 0.25&  1.95 & 7.7 \\
F1000W & 2.91 & 0.32& 1.8 &  10.0 \\
F1130W & 3.27 & 0.36& 0.73 & 11.3 \\ 
F2100W & 6.09 & 0.67& 4.58 & 21.0 \\

\bottomrule[1pt] 
\bottomrule[1pt] 
\label{tab:FWHM}
\end{tabularx}
\end{table}
 
\subsection{The Fourier transform, power spectrum and noise} \label{sec:method:FourierTransf}

\begin{figure}[t!]
\centering
\includegraphics[width=1\columnwidth]{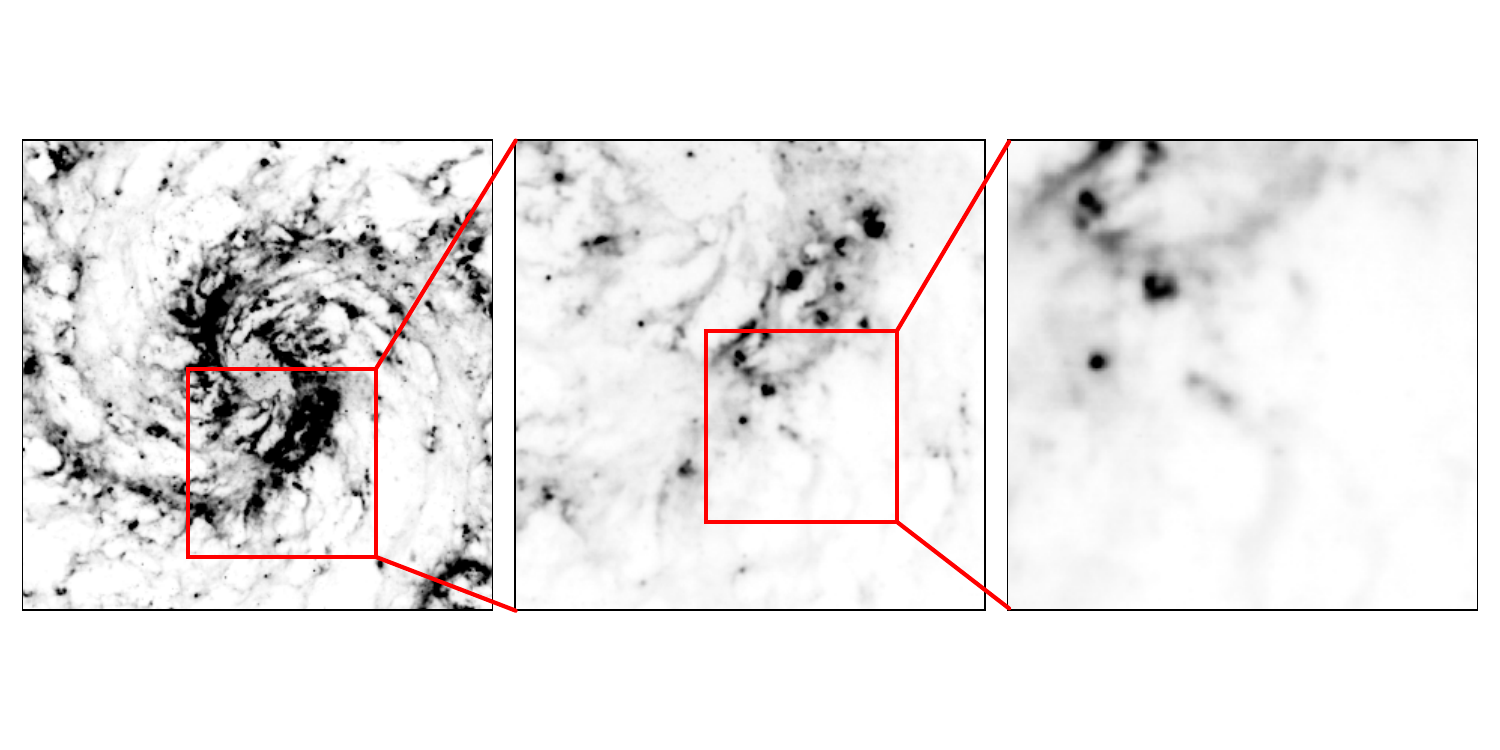}
\caption{ NGC 628 in F770W zoomed in at different points shown by the red square. The structure of NGC\,628 decrease gradually when going down in scale. The abundance of structure on a given scale is what is quantified by the power spectrum, and thus the slope of the power spectrum, $\alpha$, quantified the decrease in structure.}
\label{fig:zoom}
\end{figure}

The 2-D Fourier transform takes a signal, $f(x, y)$, in physical space, and transforms it to frequency space $\mathcal{F}(k_x, k_y)$. The power spectrum of a 2-D signal
\begin{equation}
    \begin{split}
    P(k_x, k_y) = |\mathcal{F}(k_x, k_y)|^2,
    \end{split}
\end{equation}
abolish phase components of the Fourier transform, and quantifies the magnitude of each frequency in $\mathcal{F}(k)$. For spatial signals, the power spectrum quantifies the amount of structure present on each spatial scale $\ell = 2\pi/k$ (see Fig.~\ref{fig:zoom}).

The 2-D power spectrum for a square image can be dimensionally reduced by binning all intensities with a constant radial wavenumber, $k_r = \sqrt{k_x^2 + k_y^2}$. The lower and upper bounds of the scales analyzed with $P(k_r)$ are then

\begin{equation}
    \begin{split}
        d_{lower} &= \text{FWHM}(\lambda) \, d_{galaxy} \\
        d_{upper} &=  \mathrm{MIRI}_{\mathrm{FOV}} \, d_{galaxy} = \frac{\text{MIRI}_{\mathrm{FOV}}}{\text{FWHM}(\lambda)} d_{lower},
    \end{split}
    \label{eq:lower_and_upper_bound}
\end{equation}

\text{FWHM} is the full-width half maximum in radians in each filter (see Table\,\ref{tab:FWHM}),  \text{MIRI}$_{FOV}$ refers to the field of view ($73.5''$) of the MIRI, and $d_{galaxy}$ is the distance to the galaxy. A 1-D power spectrum is constructed by plotting the median of these bins, using the 16th and 84th percentiles as a measure of the spread and Poisson uncertainties as the uncertainty on the median. This displays the overall distribution of the structure on any scale.

When the power spectrum of the Fourier domain is well-described by a power law, 
\begin{equation}
    P(k_r) = bk_r^{-\alpha}
    \label{eq:powerlaw}
\end{equation}
the slope, $\alpha$, of the power spectrum characterize the self-similar parts of the power spectrum. We note that other characteristics of the power spectrum, such as break points, contain other information. To include a break point at a scale $\ell_0 = 2\pi/k_0$, we describe the power law as a piecewise function,

\begin{equation}
    P(k_r) \begin{cases} b\cdot k_r^{-\alpha_1} & \text{if } k_r < k_0 \\
    b \cdot k_0^{\alpha_2-\alpha_1} \cdot k_r^{-\alpha_2}   & \text{if } k_r > k_0  
        \end{cases}
    \label{eq:broken_powerlaw}
\end{equation}

where $b \cdot (k_0)^{\alpha_2-\alpha_1}$ ensures continuity between the two power laws. Physically, the slope of the power spectrum quantifies how the amount of structure decreases when decreasing in scales. A steep slope corresponds to less small-scale structure as compared to a shallower slope (see representation in physical space in Fig. ~\ref{fig:zoom}).

The Fourier transform, being a correlation function, is generally sensitive to systematic effects, as it can amplify artificial scales and patterns in the signal that may be introduced by instrumental or observational artifacts. In this analysis, such effects have either been corrected if possible, or the portion of the power spectrum where the systematic effect dominates has been excluded.
The resolution of the imaging is described as a convolution between the image and a Gaussian, $g$, characterized by the full width at half maximum (FWHM) of the instrument's point spread function (PSF). This convolution impacts the power spectrum at all scales, but its effect can be removed using the convolution theorem, since the FWHM of JWST imaging is well characterized (see Table~\ref{tab:FWHM}).
White noise begins to dominate in the Fourier domain at around a scale of $\ell \approx 2\text{FWHM}$ for all filters. This manifests as a flattening of the power spectrum, consistent with the spectral shape of white noise. Accordingly, the power spectrum is truncated at this scale. Note, that this is simply the scale on which the PSF of the telescope begins to smear out the structure in the observation.

Foreground stars play no significant role in the power spectrum within the MIR bands, as the local abundance of red dwarfs, whose emission peaks $\sim$2$\mu$m, contributes minimally in this wavelength regime. However, the impact of foreground stars on the power spectrum is substantial in the near-infrared (NIR), which is why NIR filters are excluded from this study.

\section{Results}\label{sec:result}

The method described in Sect.~\ref{sec:Method} is used to find the power spectrum of 13 galaxies from the PHANGS-JWST treasury for each of the following MIRI filters; F770W, F1000W, F1130W and F2100W. We note this galaxy sample includes objects with bright luminous cores (ie. NGC\,1300, NGC\,1433, NGC\,1512, NGC\,1672, NGC\,3351, NGC\,4303 and NGC\,4321) and more extended flocculent morphologies (ie. NGC\,628, NGC\,1087, NGC\,1512, NGC\,2835, NGC\,4254 and IC\,5332). 
Each 1D power spectrum is the median of the 2D power spectrum at their respective wavenumber, $k_r$, and is plotted for NGC\,628 in Fig.~\ref{fig:powerspektre}. The power spectra of the remaining 12 galaxies is displayed in the Appendix,~\ref{sec:Appendix_powerspectra}.

\begin{figure*}
\centering
\includegraphics[width=1\linewidth]{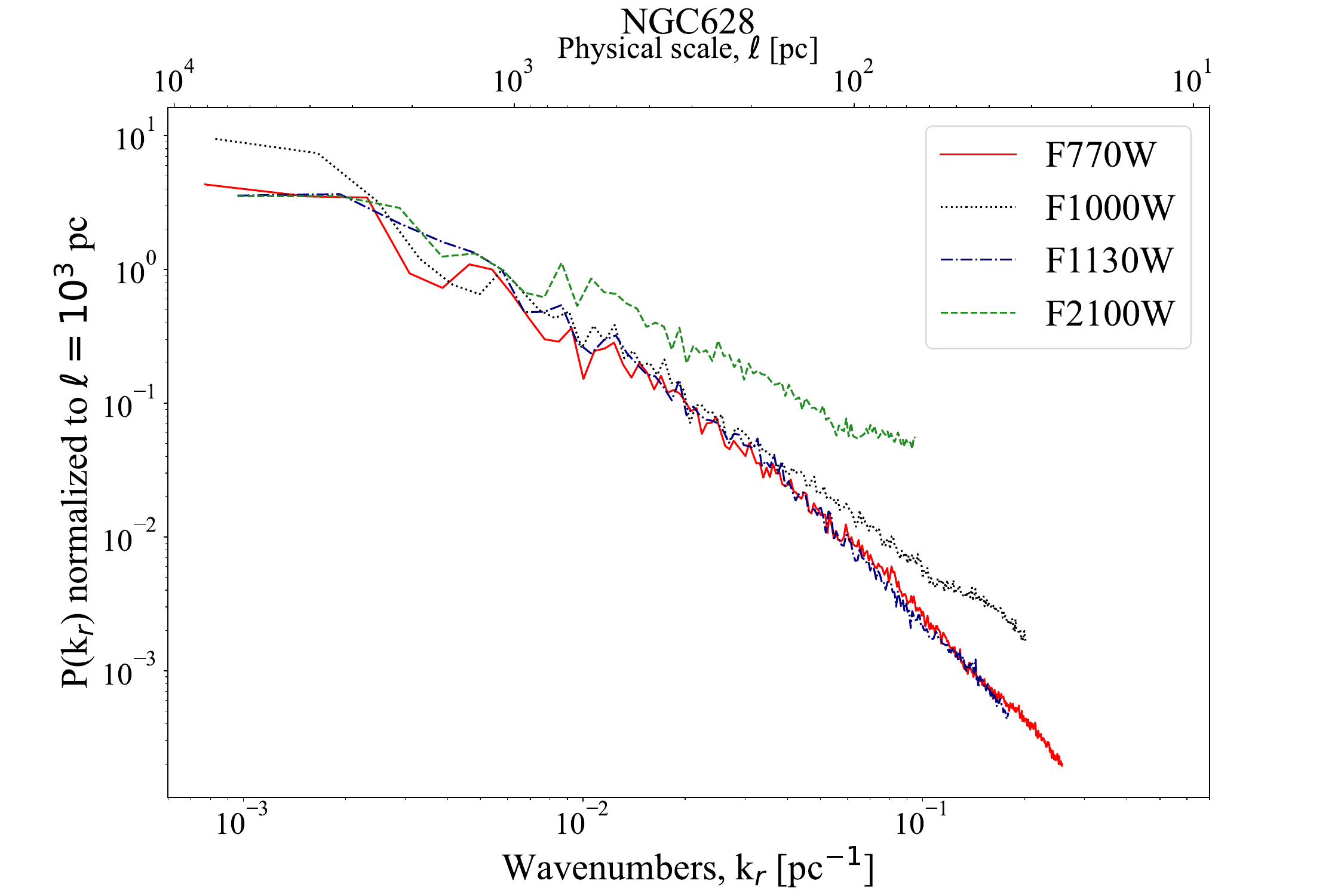}
\caption{ Power spectra for NGC 628 in the 4 MIRI filters, F770W (fully drawn), F1000W (dotted), F1130W (dashdotted), and F2100W (dashed), processed as described in Sect.~\ref{sec:Method}. The wavenumber relates to the physical scale $\ell = 2\pi/k$.
The power spectrum of the PAH filters, F770W and F1130W, are more alike than the power spectrum for F1000W and F21000W. This indicates that F770W and F1130W correspond to structures that are alike, as expected from the tracing of PAH in these filters. The steeper slope of the PAH filters indicate that there's relatively less small-scale PAH structure than dust in NGC\,628.}
\label{fig:powerspektre}
\end{figure*}

\begin{table}[h!]
\caption{The median slope of 13 galaxies in four MIRI filters fit for all scales with the 16th and 84th percentile, quantifying the distributions of slopes. The slope for a broken up power law above and below the break scale $\ell_0$ are shown in the Appendix, Tables \ref{tab:F770WandF1000W_slopes} and \ref{tab:F1130WandF2100W_slopes}. For the global upper and lower scales see Eq. \ref{eq:lower_and_upper_bound}.
$\alpha_{F770W}$ and $\alpha_{F1130W}$, the PAH-dominated bands, are generally larger than $\alpha_{F1000W}$ and $\alpha_{F2100W}$, the dust-dominated bands. This indicate, that compared to dust, PAH structure is more suppressed on small scales. }
\label{tab:slopes_all_k}
\setlength{\tabcolsep}{4pt} 
\begin{tabularx}{1\columnwidth}{lllll}
\toprule[1pt] 
\toprule[1pt] 
Galaxy &  $\alpha_{F770W}$ & $\alpha_{F1000W}$ & $\alpha_{F1130W}$ & $\alpha_{F2100W}$\\
\midrule 
NGC 628 & 2.12$\pm$0.04 & 1.71$\pm$0.08 & 2.21$\pm$0.07 & 1.19$\pm$0.15  \\
NGC 1087 & 2.37$\pm$0.07 & 2.14$\pm$0.09 & 2.27$\pm$0.54 & 1.33$\pm$0.16 \\ 
NGC 1300 & 2.2$\pm$0.07 & 1.19$\pm$0.08 & 1.64$\pm$0.11 & 0.9 $\pm$ 0.4 \\ 
NGC 1433 & 2.30$\pm$0.06 & 2.56$\pm$0.05 & 2.18$\pm$0.07 & 1.2$\pm$0.1 \\ 
NGC 1512 & 2.11$\pm$0.05& 1.69$\pm$0.08 & 1.95$\pm$0.08 & 1.3$\pm$0.11 \\ 
NGC 1672 & 2.21$\pm$0.11 & 1.3$\pm$0.3 & 2.0$\pm$0.9 & 1.05$\pm$0.11 \\ 
NGC 2835 & 1.94$\pm$0.07 & 1.30$\pm$0.08 & 1.61$\pm$0.09 & 0.73$\pm$0.11 \\ 
NGC 3351 & 2.47$\pm$0.06 & 1.76$\pm$0.09 & 2.17$\pm$0.18 & 1.04$\pm$0.12 \\
NGC 4303 & 2.34$\pm$0.06 & 1.2$\pm$0.2 & 2.2$\pm$0.8 & 1.6$\pm$0.2 \\
NGC 4254 & 2.20$\pm$0.07 & 2.04$\pm$0.10 & 2.13$\pm$0.13 & 1.4$\pm$0.2 \\
NGC 4321 & 2.06$\pm$0.08 & 0.92$\pm$0.12 & 1.4$\pm$0.8 & 0.9$\pm$0.2 \\
NGC 5068 & 2.08$\pm$0.07 & 1.47$\pm$0.09 & 1.85$\pm$0.09 & 0.75 $\pm$ 0.12 \\ 
IC 5332 & 1.58$\pm$0.06 & 0.84$\pm$0.07 & 1.53$\pm$0.09 & 0.58$\pm$0.15 \\
\midrule
Median & 2.19 $^{+0.16}_{-0.17}$ & 1.47 $^{+0.57}_{-0.31}$  & 1.99 $^{+0.19}_{-0.35}$ & 1.05$^{+0.28}_{-0.31}$ \\
\bottomrule[1pt]
\bottomrule[1pt]
\end{tabularx}
\end{table}

\begin{table}[h!]
\caption{Median p-value between slopes for all $\ell$ for each pair of filters (see Table~\ref{tab:slopes_all_k}) within the same galaxy. As expected from tracing, $\alpha_{F770W}$ and $\alpha_{F1130W}$, the two PAH bands, are similar, while the dust-dominated bands $\alpha_{F1000W}$ and $\alpha_{F2100}$ also are similar.}
\centering
\begin{tabular}{l|llll}
\toprule[1pt]
\toprule[1pt]
       & F770W & F1000W & F1130W & F2100W \\ \hline
F770W  & 1     & $1.8 \cdot 10^{-12}$    & $3.4 \cdot 10^{-4}$   & $6.4 \cdot 10^{-18}$   \\
F1000W & $9.8 \cdot 10^{-13}$  & 1      & $5.7 \cdot 10^{-6}$   & $9.0\cdot10^{-3}$   \\
F1130W & $3.4 \cdot 10^{-4}$ & $5.7 \cdot 10^{-6}$   & 1      & $4.9 \cdot 10^{-9}$ \\
F2100W & $9.8 \cdot 10^{-18}$   & $9.0\cdot10^{-3}$   & $4.9 \cdot 10^{-9}$   & 1      \\ 
\bottomrule[1pt]
\bottomrule[1pt]
\end{tabular}
\label{tab:KS_test}
\end{table}

\begin{figure*}[t!]
    \centering
    \includegraphics[width=1\linewidth]{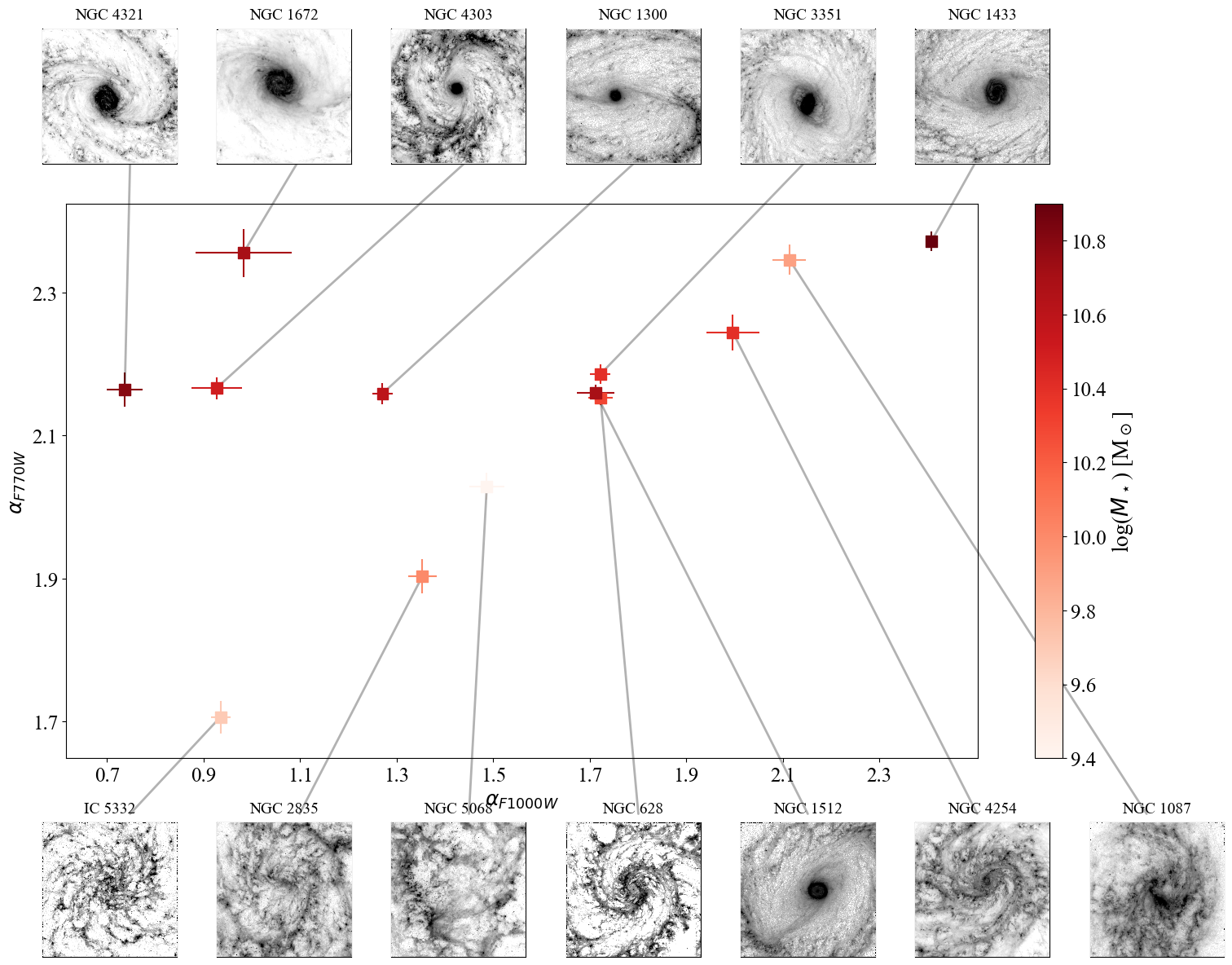}
    
    \caption{$\alpha_{F1000W}$ against $\alpha_{F770W}$ with the colorbar of the marker indicating the stellar mass of the galaxy. The galaxy yielding each point is shown in a cut out in F770W. A similar trend between slopes are found between all other filters. There's an indication of two distinct trends, where galaxies with very luminous bulges lie on the upper trend, while more equally distributed galaxies lie on a lower trend. Furthermore, $\alpha_{F1000W}$ are spread over a larger range of values than $\alpha_{F770W}$. This is expected to be due to F1000W tracing a larger mix of warm dust and PAHs.}
    
    \label{fig:F1000WvsF770W_correlation}
\end{figure*}

\begin{table}[t!]
\caption{The median slopes for power law fits for large scales ($\ell > \ell_0$) and small scales ($\ell < \ell_0$) (shown in the Appendix, Tables \ref{tab:F770WandF1000W_slopes} and \ref{tab:F1130WandF2100W_slopes}), and the break scale, $\ell_0$, for the broken power law shown with the 16th and 84th percentile showing the distribution of slopes in a given filter across the entire galaxy sample. For the global upper and lower scales see Eq. \ref{eq:lower_and_upper_bound}.}

\begin{tabularx}{1\columnwidth}{lllll}
\toprule[1pt] 
\toprule[1pt] 
Filter & All $\ell$ & $\ell > \ell_0$ & $\ell < \ell_0$ & $\ell_0$ \\

\textbf{$\alpha_{\mathrm{F770W}}$} & 2.16$^{+0.16}_{-0.15}$ & 1.60 $^{+0.23}_{-0.27}$ &2.45 $^{+0.13}_{-0.24}$ &  235$^{+590}_{-140}$ pc\\

\textbf{$\alpha_{\mathrm{F1000W}}$} & 1.48$^{+0.33}_{-0.47} $ & 1.66$^{+0.33}_{-0.76}$ & 1.45$^{+0.45}_{-0.64}$ &  103$^{+580}_{-63}$ pc\\

\textbf{$\alpha_{\mathrm{F1130W}}$} & 1.88 $^{+0.25}_{-0.37} $ & 1.70$^{+0.29}_{-0.48}$ & 1.89$^{+0.73}_{-0.27}$ &  157$^{+540}_{-90}$ pc \\

\textbf{$\alpha_{\mathrm{F2100W}}$} & 0.94 $^{+0.23}_{-0.28} $ & 1.12$^{+0.06}_{-0.60}$ &  0.82$^{+0.46}_{-0.35}$  &  130$^{+310}_{-90}$ pc\\

\bottomrule[1pt]
\bottomrule[1pt]

\end{tabularx}
\label{tab:median_small_and_large_slopes}
\end{table}

The power spectrum for all the 13 galaxies in the four MIRI filters, F770W, F1000W, F1130W and F2100W are well described by one or two power laws, almost three decades of spatial scale ($10^4-20$)\,pc for F770W. This is remarkable considering the large diversity of galactic morphologies studied (see Fig.~\ref{fig:F1000WvsF770W_correlation}). The galaxies with a bright galactic core display periodic bumps in the power spectrum and are the worst fit to a power law.

\begin{figure*}
    \centering
    \includegraphics[width=1\linewidth]{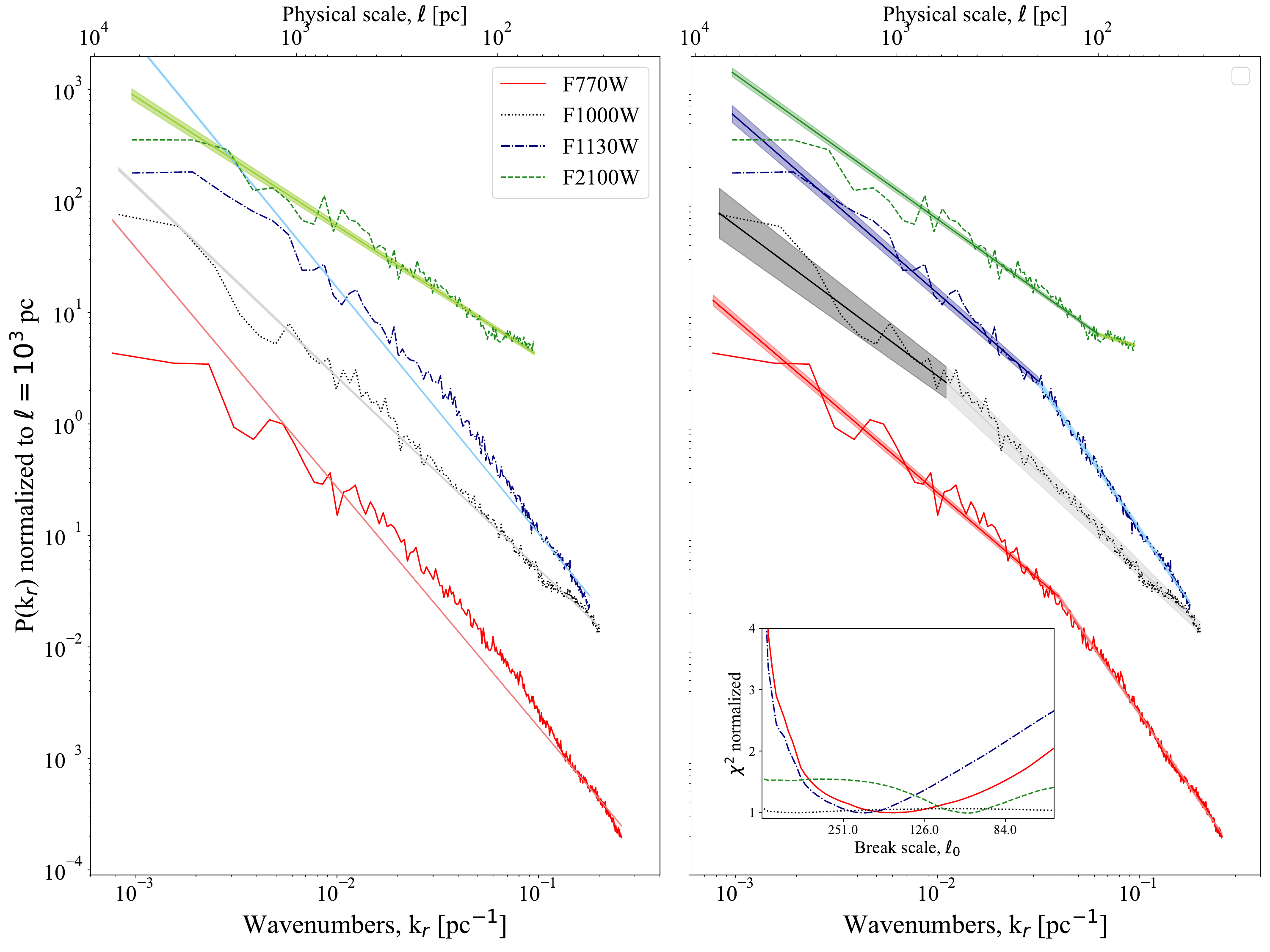}
    \caption{Power spectrum of NGC\,628 fit to a single power law (left) and a broken power law (right) with the shading being the uncertainty of the fit. The $\chi^2$-landscape as a function of the break scale, $\ell_0$, is included as an inset for the broken power law (right).  It's clear, that the power law fit of the PAH-dominated bands, F770W and F1130W, becomes a dramatically better when fit to a broken power law from the strong minima in the $\chi^2$-landscape as a function of $\ell_0$. In contrast, this is not the case for the dust-dominated band, F1000W. We note a small minimum for F2100W, but with an insignificant change in $\chi^2$. }
    \label{fig:fit_of_powerlaws}
\end{figure*}

\subsection{ $\alpha$-distribution across galaxy sample} \label{subsec:alpha_dist}
The distribution of $\alpha$ for each filter is quantified through the median and the 16th and 84th percentiles. 
The PAH-dominated filters, F770W and F1130W, contain around a magnitude more PAH emission than the dust-dominated filters, F1000W and F1130W. F770W is the strongest PAH band, as F1130W contains about half the PAH emission of F770W \citep{Hensley_2023}. Other studies of PHANGS-JWST imaging find that F1000W mirrors the PAH bands \citep{Leroy_2023}, though the specific nature of why this is so is uncertain. For this reason, we use F770W as the strongest PAH band and F2100W as the strongest dust band. The median slopes of these bands for the total ensemble of galaxies for all $\ell$ is $\alpha_{\mathrm{F770W}} = 2.16 ^{+0.16}_{-0.15} $ and $\alpha_{\mathrm{F2100W}}= 0.94 ^{+0.23}_{-0.28}$. The slopes of the remaining filters, which trace different fractions of dust and PAH, lie somewhere between these two extremes (see Table \ref{tab:median_small_and_large_slopes}). 
The large scale $\alpha$, $\ell>\ell_0$, of all bands except F2100W are very similar, thus, thus gradual change of the slope for changing fractions of PAH is primarily driven by the steep slope for the PAH bands on small scales. 

\subsection{ Comparing PAH and thermal dust bands} \label{subsec:comparing_PAH_and_dust}
Based on the determined slopes of the power spectra, $\alpha$, (see Table~\ref{tab:slopes_all_k}), the highest degree of similarity is found between the dust-dominated bands, F1000W and F2100W, and the PAH-dominated bands, F770W and F1130W, based on the median p-value between $\alpha$ for all pairs of filters within the same galaxy in Table~\ref{tab:KS_test}. This is as expected from the similar tracing of the two pairs of filters, and it shows that the power spectrum preserves the morphological properties of the constituents traced in the specific filters. The relatively large similarity between F1000W and F1130W can be explained by the larger superposition of dust and PAH emission in F1130W compared to F770W.

In Figure \ref{fig:F1000WvsF770W_correlation}, $\alpha_{F770W}$ is plotted against $\alpha_{F1000W}$ with the color indicating the stellar mass. Firstly, this plot reveal a clear separation in $\alpha$ for galaxies with bright galactic cores (clustered near the top of the figure, except NGC\,1512). These galaxies show a nearly constant $\alpha\approx2.2$. For the remaining galaxies, $\alpha_{F770W}$ and $\alpha_{F1000W}$ are strongly correlated Stellar mass is used as a proxy for galactic properties, though no strong trend is observed. The slopes across all $\ell$ are consistently lower for dust bands than for PAH bands, indicating less small-scale PAH structure relative to dust. This suggests a fundamental difference in the spatial distribution of these constituents.

\subsection{ Broken power law for PAH emission }

\begin{figure*}[t!]
\centering
\includegraphics[width=1.99\columnwidth]{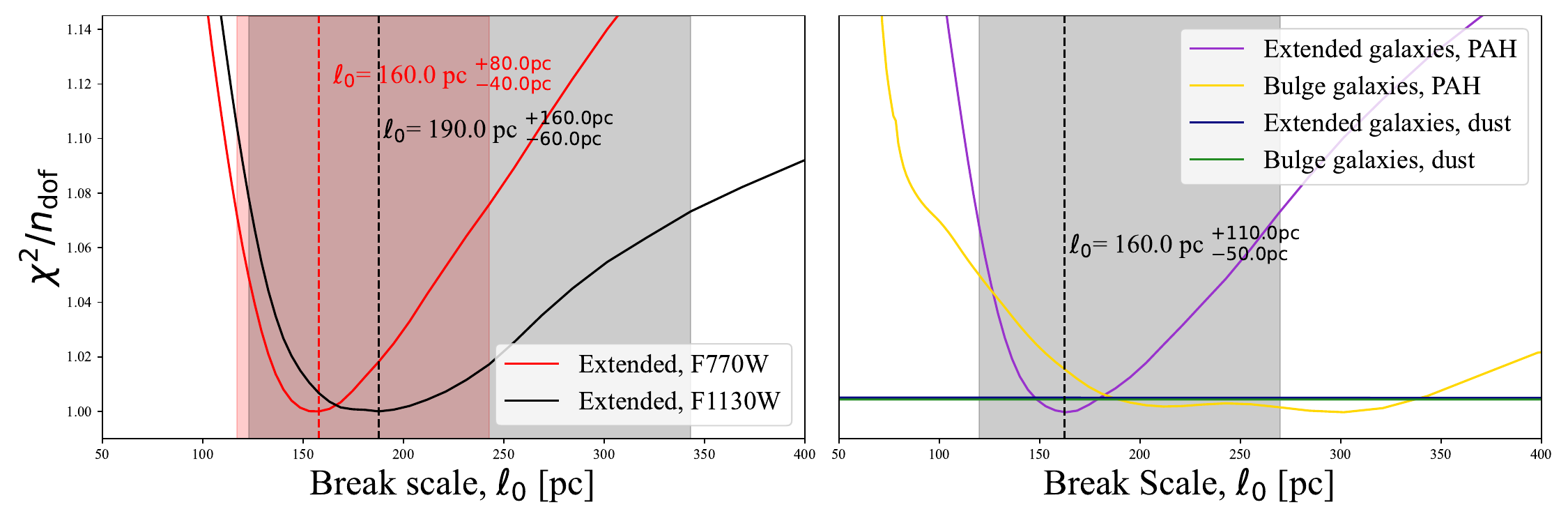}
\caption{ $\chi^2$-distribution with errors normalized at $\min(\chi^2(\ell))$ for a combined broken power law fit with a single break scale, $\ell_0$, for the ensemble of galaxies with extended morphologies (ie. NGC\,628, NGC\,1087, NGC\,1512, NGC\,2835, NGC\,4254 and IC\,5332) for F770W and F1130W separately (left panel) and combining PAH filters for extended and bulge galaxies (ie. NGC\,1300, NGC\,1433, NGC\,1512, NGC\,1672, NGC\,3351, NGC\,4303 and NGC\,4321), along with the extended and bulge galaxies in the dust filters (right panel). From this, it's clear that the break-scale for F770W and F1130W are consistent with each other. 
The structure of the PAH filters have a collective break scale, $\ell_0 = 160 \mathrm{pc}^{+110\mathrm{pc}}_{-50\mathrm{pc}}$, while the  structure within dust filters have none. We note that the constrain of the bulge galaxies are worse, but still consistent with no break in dust-bands and the break-scale for PAH-bands. }
\label{fig:p_of_chi_square}
\end{figure*}

Fitting a broken power law, segmented at a scale $\ell_0$, improves the $\chi^2$ for the PAH-dominated bands (F770W and F1130W) by a factor of $\sim$2, while the $\chi^2$ of dust-dominated bands (F1000W and F2100W) remains unchanged (see Figure \ref{fig:p_of_chi_square}). We caution that the statistical uncertainty in $P(k)$ derived on various scales should perhaps not be considered entirely independent. That is particularly acute for small scale modes, which are increasingly closer approximations of each other and approximately derive from the same structures in the image. Thus, the uncertainty in $P(k)$ at various scales, $k$, must be correlated, making the direct interpretation of the exact $\chi^2$ tenuous. Nevertheless, given the several-fold increase in the goodness of fit, we argue that $\chi^2$ remains a useful statistical measure. We compute the uncertainty on the break-scale $\ell_0$ at the 1$\sigma$ level given the the $\chi^2$ distribution, but with errors normalized at $\min(\chi^2(\ell))$ to alleviate the potential correlation between different scale.

The landscape of $\chi^2(\ell_0)$ (Figure~\ref{fig:fit_of_powerlaws}) differs markedly between the PAH and dust bands. The PAH bands exhibit a clear preference for a break scale, while the dust bands show none. Thus, the break scale $\ell_0$ inferred from dust-dominated bands is statistically unconstrained, indicating that deviations from a single power-law model are not required by the data. Galaxies with compact, luminous cores show complex, multi-peaked power spectra, resulting in broader distributions in $\ell_0$ (see Table~\ref{tab:median_small_and_large_slopes}).

In Figure \ref{fig:p_of_chi_square} (left panel), the $\chi^2$-landscape for the break scale, $\ell_0$ for a ensemble fit of all galaxies with extended morphologies (ie. NGC\,628, NGC\,1087, NGC\,1512, NGC\,2835, NGC\,4254 and IC\,5332) are shown for F770W and F1130W individually. The break scale of the two filters, $\ell_0=160\mathrm{pc}^{+80\mathrm{pc}}_{-40\mathrm{pc}}$ and $\ell_0=190 \mathrm{pc}^{+160\mathrm{pc}}_{-60\mathrm{pc}}$, respectively, are consistent within 1$\sigma$. Furthermore, an ensemble fit for both PAH bands for the sample of extended and bulge morphology galaxies separately (see Figure \ref{fig:p_of_chi_square}, right panel). The bulge galaxies are not as tightly constrained, due to the strong oscillations induced in the power spectrum from the modes of the luminous bulge. Therefore, we interpret the break scale of the extended galaxies as the best representation of the true break scale for PAH structure, $\ell_0 = 160 \mathrm{pc}^{+110\mathrm{pc}}_{-50\mathrm{pc}}$.

\section{Discussion} \label{sec:discuss}

Across galaxies as seen in Fig. \ref{fig:F1000WvsF770W_correlation}, the distribution of $\alpha$ for the PAH-dominated band is narrower than for dust-dominated bands by a factor of 2-3, especially prominent for large scales ($\ell > \ell_0$). This indicates that the PAH structures are more homogeneous across galaxies than hot dust. While PAH specifically trace star formation and photo-dissociation regions, warm dust traces the overall stellar population (and thus the integrated star formation across time). This indicates a diversity in the older galactic structures between galaxies and a homogeneity of the traced younger stellar population. The homogeneity of PAH structure could be due to photo-dissociation regions being influenced by strong feedback effects, which are similar across local galaxies, because the driving sources of energy, through star formation, are identical. In contrast, for this interpretation, the total stellar population would have variability due to the diversity of environmental histories. For instance, we note that the two main galaxy types analyzed, e.g. bright nucleus versus more extended structures, exhibit different trends (see Fig. \ref{fig:F1000WvsF770W_correlation}). The bright bulges, which share a dominant star-formation in their core and exhibit a single power law slope, $\alpha\sim2.2$, in the PAH-bands. While the slope of the extended galaxies have a strong correlation between $\alpha_{\mathrm{F770W}}$ and $\alpha_{\mathrm{F1000W}}$.

The broken power law is a better fit for the PAH-dominated bands, while there is no significant improvement for the dust-dominated bands (see Figure \ref{fig:fit_of_powerlaws}). A break in the power spectrum implies a characteristic physical scale and is typically related to a transition between regimes, e.g. the dimensionality of turbulence or energy injection at a specific scale.
Thus, the strong preference for a break in the power spectrum by the PAH-dominated bands, show that PAH structure, unlike dust structure, has a characteristic scale, $\ell_0 = 160 \mathrm{pc}^{+110\mathrm{pc}}_{-50\mathrm{pc}}$, 
beneath which, the PAH emission structures are suppressed. This is perhaps natural as PAH emission trace young stars in the surrounding photo-dissociation regions and is therefore prone to characteristic scales whether from SNe feedback or spatial extent of star formation (e.g. the Jeans length). 

The characteristic scale, $\ell_0$, of the PAH structures is consistent with the size of Giant Molecular Clouds $\sim100\mathrm{pc}$ \citep{Murray_2011}, and in agreement with the scale of $\sim100\mathrm{pc}$ found by \citep{2021_Pessa} as the tightest relation for the molecular gas main sequence for the PHANGS galaxies. Thus, it seems to resemble a characteristic size of star formation. For this interpretation, the break scale of PAH structure could originate from small-scale PAH structures, $\ell < \ell_0$, being dominated by stellar feedback effects and strong UV radiation.

As the break only occurs for the PAH-dominated filters, it is likely not related to the scale height of the disk, where different spatial scales are characterized by 2-D or 3-D turbulence, as has been suggested by previous studies \citep{Block2009}. Furthermore, \cite{Meidt_2023} determine a turbulent Jeans length $\lambda_J\approx (400-1150)\mathrm{pc}$ and a filament spacing $\lambda_{\mathrm{fil}} \approx (400-1100)\mathrm{pc}$ across four JWST galaxies (NGC\,628, IC\,5332, NGC\,1365 and NGC\,7496) and different radial regimes of the galaxies. These characteristic scales remains much larger than that suppressing PAH structures found in this study.

NGC\,628 and NGC\,3351 has previously been studied using the $\Delta$-variance spectrum of HI gas by \cite{2021_Dib}, where they despite the different proxy find a characteristic scale of $\sim$320 pc and $\sim$243 pc, respectively. The first of these scales is largely in tension with the characteristic scale determined for PAH structures for NGC\,628, while it is consistent with the largely ill-constrained break-value for NGC\,3351 (see \ref{sec:Appendix_slopes}). Furthermore, in addition to characteristic scales around $\sim10^2 \mathrm{pc}$, \cite{2021_Dib} determines scales on the order of a few to several kpc. The latter characteristic scales of HI gas are much larger than those determined for PAH structures in this work, and are thus most likely unrelated to them.

Ultimately, these results highlight how the capabilities of the JWST telescope has opened a new diagnostic regime, beyond the resolution of previous near- and mid-infrared analysis and the scope of single power law power spectra. In this new power spectral regime spanning orders-of magnitude and down to the tens of parsec scale, different bands (and the constituents therein traced) instead express a diverse set of signatures, which hint at the unique physics in connecting scales, forming and suppressing structures. Unraveling these complex patterns between the various scales of dust and PAH structures may finally permit a deeper understanding of these constituents and the physical environments they embed.  \newline

\section{Acknowledgments}
The authors would like to express gratitude to Chris Hayward for discussions and comparisons to simulations, to Aigen Li for insight on PAHs, to Connor McPartland and Erik Høg for discussion of the effect of foreground stars, to the anonymous referee for insightfull feedback and comments. The Cosmic Dawn Center (DAWN) is funded by the Danish National Research Foundation under grant DNRF140. AS is co-funded by the European Union (ERC, HEAVYMETAL, 101071865). Views and opinions expressed are, however, those of the authors only and do not necessarily reflect those of the European Union or the European Research Council. Neither the European Union nor the granting authority can be held responsible for them. 


\bibliographystyle{aasjournal}
\bibliography{main}

\newpage

\section{Appendix}
\subsection{Power spectra for all galaxies}\label{sec:Appendix_powerspectra}

\begin{figure*}[h!]
\centering
\includegraphics[width=0.85\linewidth]{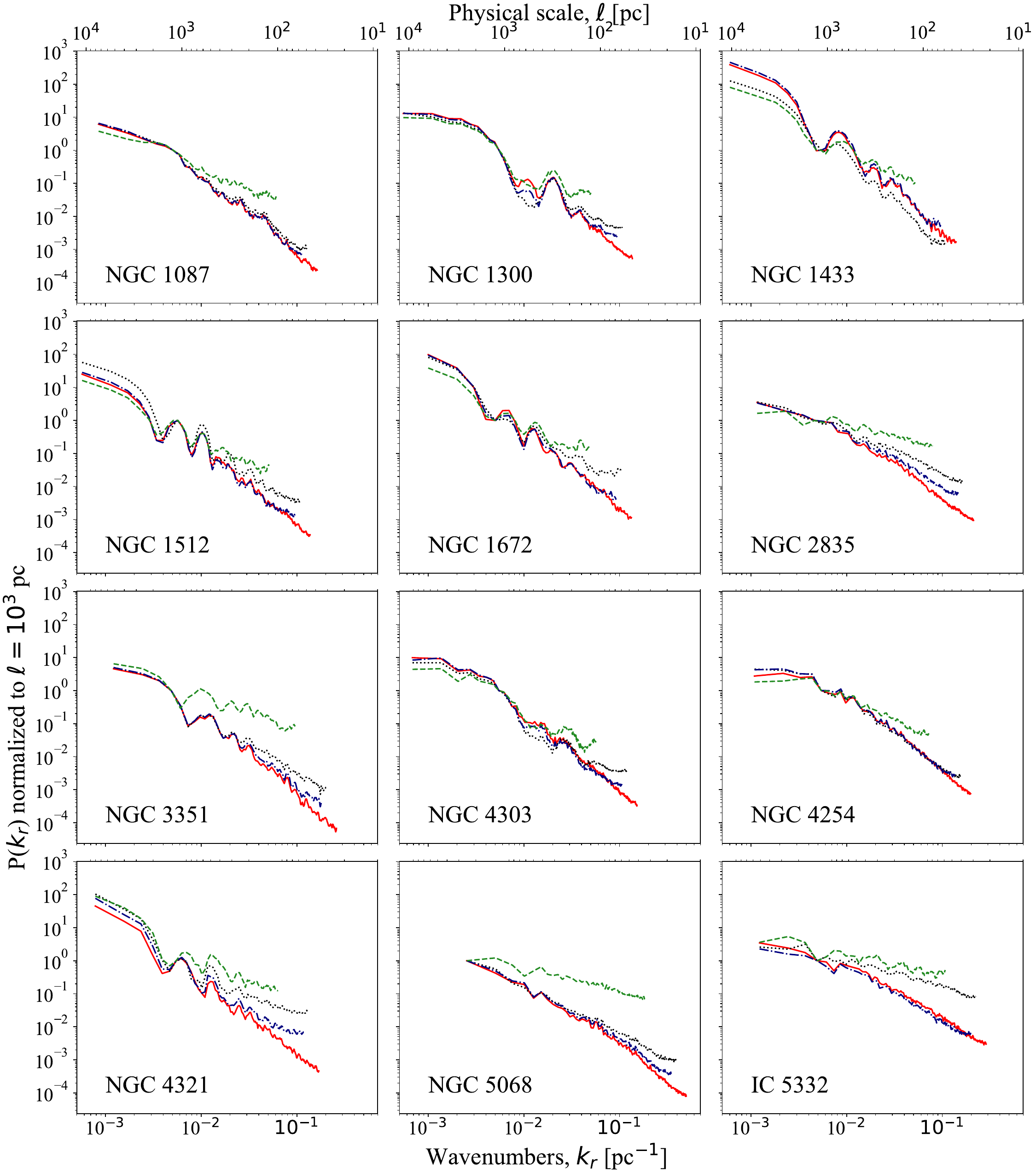}
\caption{Power spectra for 12 PHANGS-JWST galaxies in the four MIRI filters, F770W (red, fully drawn), F1000W (black, dotted), F1130W (blue, dashdotted), and F2100W (green, dashed). Note the periodic bump structures in the spectra of NGC\,1300, NGC\,1433, NGC\,1512, NGC\,1672, NGC\,3351, NGC\,4303 and NGC\,4321 correlates perfectly with the galaxies with a very luminous galactic core, as this introduce a prominent mode.}
\label{fig:all_powerspectra}
\end{figure*}

\newpage
\subsection{Slope for power spectra for all galaxies}\label{sec:Appendix_slopes}

\begin{table*}[h!]
\centering
\setlength{\tabcolsep}{4pt} 
\begin{tabularx}{\textwidth}{l|llll|llll}
\toprule[1pt] 
\toprule[1pt] 

&       & $\alpha_{F770W}$ &     &       & $\alpha_{F1000W}$   \\
& all $\ell$ & $\ell > \ell_0$ & $\ell < \ell_0$ & $\ell_0$ [pc] & all $\ell$ & $\ell > \ell_0$ & $\ell < \ell_0$ & $\ell_0$ [pc] \\

\midrule 
NGC 628  & 2.16 $\pm$ 0.04 & 1.36 $\pm$ 0.02  & 2.59 $\pm$ 0.01 & $150^{+35}_{-27} $ & 1.72 $\pm$ 0.06 & 1.63 $\pm$ 0.01  & 1.73$\pm$0.01 & $60^{+3}_{-3}$ \\

NGC 1087 & 2.32 $\pm$ 0.05 & 2.02 $\pm$ 0.01 & 2.67$\pm$0.01 & $150^{+68}_{-27}$ & 2.09 $\pm$ 0.07 & 1.93 $\pm$ 0.01  & 2.33 $\pm$ 0.01 & $60^{+3}_{-3}$ \\

NGC 1300 & 2.14 $\pm$ 0.04 & 0.87 $\pm$ 0.08  & 2.25 $\pm$ 0.02 & $300^{+67}_{-81}$ & 1.05 $\pm$ 0.09 & -3.5 $\pm$ 0.21  & 1.31 $\pm$ 0.01 & $370^{+64}_{-100}$ \\

NGC 1433 & 2.37 $\pm$ 0.03 & 1.85 $\pm$ 0.02 & 2.65 $\pm$ 0.01 & $840^{+202}_{-705}$ & 2.53 $\pm$ 0.03 & 2.36 $\pm$ 0.01  & 2.68 $\pm$ 0.01 & $720^{+242}_{-190}$ \\

NGC 1512 & 2.23 $\pm$ 0.02 & 1.79 $\pm$ 0.01 & 2.44 $\pm$ 0.01 & $570^{+78}_{-243}$ & 1.54 $\pm$ 0.02 & 1.70 $\pm$ 0.01  & 1.47 $\pm$ 0.01 & $100^{+46}_{-27}$ \\

NGC 1672 & 2.31 $\pm$ 0.08 & 1.67 $\pm$ 0.04 & 2.57 $\pm$ 0.01 & $140^{+63}_{-51}$ & 1.03 $\pm$ 0.16 & 1.8 $\pm$ 0.03  & 0.74 $\pm$ 0.02 & $100^{+7}_{-18}$ \\

NGC 2835 & 1.93 $\pm$ 0.06 & 1.45 $\pm$ 0.02 & 2.10 $\pm$0.01 & $250^{+191}_{-62}$ & 1.33 $\pm$ 0.07 & 0.96 $\pm$ 0.04  & 1.43$\pm$0.01 & $180^{+68}_{-59}$ \\

NGC 3351 & 2.42 $\pm$ 0.03 & 1.73 $\pm$ 0.03 & 2.63 $\pm$ 0.01 & $300^{+223}_{-128}$ & 1.70 $\pm$ 0.03 & 1.22 $\pm$ 0.03  & 1.79 $\pm$ 0.01 & $1000^{+6668}_{-1000}$ \\

NGC 4303 & 2.29 $\pm$ 0.04 & 1.85$\pm$ 0.02  & 2.47 $\pm$ 0.01 & $250^{+1333}_{-250}$ & 1.0 $\pm$0.2     & 1.96 $\pm$ 0.03    & 0.84 $\pm$ 0.01 & $80^{+14}_{-9}$ \\

NGC 4254 & 2.01 $\pm$ 0.06 & 1.49$\pm$0.01   & 2.51 $\pm$ 0.01 & $230^{+112}_{-63}$ & 1.87 $\pm$ 0.08   & 1.73 $\pm$ 0.01  & 1.97 $\pm$ 0.01 & $60^{+11}_{-8}$ \\

NGC 4321 & 2.12 $\pm$ 0.05 & 1.7$\pm$0.02   & 2.25 $\pm$ 0.01 & $180^{+163}_{-90}$ & 0.82 $\pm $ 0.13  & 1.15 $\pm$ 0.02  & 0.74 $\pm$ 0.01 & $70^{+4}_{-10}$ \\

NGC 5068 & 1.65 $\pm$ 0.04 & 0.93 $\pm$0.1   & 2.26 $\pm$ 0.01 & $90^{+24}_{-33}$ & 1.48 $\pm$ 0.07   & 1.28 $\pm$ 0.05    & 1.50 $\pm$ 0.01 & $110^{+9980}_{-8}$ \\

IC 5332  & 1.65 $\pm$ 0.05 & 1.14$\pm$ 0.02  & 1.79 $\pm$ 0.01  & $390^{+267}_{-276}$ & 0.88 $\pm$ 0.06 & 0.52 $\pm$ 0.03  & 0.95 $\pm$ 0.01 & $60^{+16}_{-5}$ \\

\midrule
Median & 2.16  & 1.60 & 2.45 & 235 pc & 1.48 & 1.66 &  1.45 & 103 pc \\
Percentiles &  $^{+0.16}_{-0.15}$ &  $^{+0.23}_{-0.27}$ & $^{+0.13}_{-0.24}$ & $^{+590}_{-140}  $ & $^{+0.33}_{-0.47}$ & $^{+0.33}_{-0.76}$ & $^{+0.45}_{-0.64}$ & $ ^{+580}_{-63} $\\

\bottomrule[1pt]
\bottomrule[1pt]
\end{tabularx}

\caption{Slope of power spectra for thirteen galaxies in two MIRI filters F770W and F1000W found for three intervals of scales; all scales, large scales ($\ell > \ell_0$) and small scales ($\ell < \ell_0$) together with the fit of $\ell_0$. The percentiles refer to the 16th and 84th percentile.}

\label{tab:F770WandF1000W_slopes}

\end{table*}

\begin{table*}[h!]
\setlength{\tabcolsep}{4pt} 
\begin{tabularx}{\textwidth}{l|llll|llll}
\toprule[1pt] 
\toprule[1pt] 

&       & $\alpha_{F1130W}$ &    &       & $\alpha_{F2100W}$   \\
& all $\ell$ & $\ell > \ell_0$ & $\ell < \ell_0$ & $\ell_0$ [pc] & all $\ell$ & $\ell > \ell_0$ & $\ell <\ell_0$ & $\ell_0$ [pc] \\

\midrule 
NGC 628  & 2.12 $\pm$ 0.06 & 1.51 $\pm$ 0.02 & 2.63 $\pm$ 0.01 & $200^{+48}_{-32} $ & 1.11 $\pm$ 0.13  & 1.13 $\pm$ 0.01  & 1.09 $\pm$ 0.02 & $100^{+13}_{-9}$ \\

NGC 1087 & 2.21 $\pm$ 0.07 & 2.00 $\pm$ 0.01 & 2.63 $\pm$ 0.02 & $150^{+44}_{-18}$ & 1.08$\pm$ 0.10     & 1.12 $\pm$ 0.02  & 0.77 $\pm$ 0.1 & $170^{+301}_{-26}$ \\

NGC 1300 & 1.50 $\pm$ 0.08 & 0.86 $\pm$ 0.11  & 1.55 $\pm$ 0.01 & $90^{+9915}_{-78}$ & 0.5 $\pm$ 0.4  & 1.67 $\pm$ 0.04   & 0.23 $\pm$ 0.02 & $200^{+915}_{-37}$ \\

NGC 1433 & 2.23 $\pm$ 0.04 & 1.71 $\pm$ 0.02 & 2.73 $\pm$ 0.02 & $840^{+140}_{-103} $ & 1.17 $\pm$ 0.05  & 1.13 $\pm$ 0.02  & 1.28 $\pm$ 0.04 & $1000^{+8572}_{-659}$ \\

NGC 1512 & 1.81 $\pm$ 0.02 & 1.98 $\pm$ 0.01 & 1.71 $\pm$ 0.01 & $100^{+34}_{-16}$ & 1.15 $\pm$ 0.04  & 1.16 $\pm$ 0.02  & 1.11 $\pm$ 0.04 & $130^{+29}_{-124}$ \\

NGC 1672 & 1.95 $\pm$ 0.13 & 2.03 $\pm$ 0.03 & 1.89 $\pm$ 0.02 & $110^{+15}_{-17}$ & 0.7 $\pm$ 0.3    & 1.17$\pm$ 0.03   & -0.27 $\pm$ 0.07 & $220^{+81}_{-34}$ \\

NGC 2835 & 1.60 $\pm$ 0.08 & 1.3 $\pm$ 0.03 & 1.73 $\pm$0.01 & $220^{+898}_{-59}$ & 0.68 $\pm$ 0.11  & 0.52$\pm$ 0.03   & 0.82 $\pm$ 0.03 & $840^{+519}_{-713}$ \\

NGC 3351 & 2.07 $\pm$ 0.03 & 1.71 $\pm$ 0.04 & 2.19 $\pm$ 0.01 & $510^{+1333}_{-408}$ & 0.94 $\pm$ 0.05  & 0.45 $\pm$ 0.04  & 1.35 $\pm$ 0.02 & $80^{+8}_{-9}$ \\

NGC 4303 & 1.85 $\pm$ 0.07 & 2.06$\pm$ 0.01  & 1.79 $\pm$ 0.01 & $80^{+27}_{-7}$ & 1.18 $\pm$ 0.11    & 1.31 $\pm$ 0.05  & 1.06 $\pm$ 0.05 & $140^{+12}_{-10}$ \\

NGC 4254 & 1.92$\pm$ 0.08  & 1.67$\pm$0.01   & 2.23$\pm$ 0.01 & $720^{+444}_{-601}$ & 1.2 $\pm$ 0.7  & 1.1 $\pm$ 0.02  & 1.43 $\pm$ 0.03 & $110^{+25}_{-15}$ \\

NGC 4321 & 1.24$\pm$ 0.2    & 1.76$\pm$0.02   & 1.08 $\pm$ 0.01 & $80^{+8}_{-9}$ & 0.83 $\pm $ 0.19 & 0.9 $\pm$ 0.02  & 0.71 $\pm$ 0.03 & $140^{+8572}_{-140} $ \\

NGC 5068 & 1.88 $\pm$ 0.06 & 1.18 $\pm$0.08    & 1.98 $\pm$ 0.01 & $100^{+36}_{-41}$ & 0.7$\pm$ 0.2   & 0.46 $\pm$ 0.06    & 0.82 $\pm$ 0.02 & $120^{+317}_{-38}$ \\

IC 5332  & 1.50 $\pm$ 0.07 & 1.23$\pm$ 0.02  & 1.63 $\pm$ 0.01  & $50^{+17}_{-5}$ & 0.5 $\pm$ 0.2  & 0.56 $\pm$ 0.02  & 0.49 $\pm$ 0.02 & $90^{+33}_{-8}$ \\

\midrule
Median & 1.88  & 1.70 & 1.89 & 157 pc & 0.94 & 1.12 & 0.82& 130 pc\\
Percentiles &  $^{+0.25}_{-0.37}$ &  $^{+0.29}_{-0.48}$ & $^{+0.73}_{-0.27}$ & $^{+540}_ {-90} $ & $^{+0.23}_{-0.28}$ & $^{+0.06}_{-0.60}$ & $^{+0.46}_{-0.35}$ & $^{+310}_{-90} $\\

\bottomrule[1pt]
\bottomrule[1pt]
\end{tabularx}

\caption{Slope of power spectra for thirteen galaxies in two MIRI filters F1130W and F2100W found for three intervals of scales; all scales, large scales ($\ell > \ell_0$) and small scales ($\ell < \ell_0$) together with the fit of $\ell_0$. The percentiles refer to the 16th and 84th percentile.}

\label{tab:F1130WandF2100W_slopes}
\end{table*}

\end{document}